\documentclass[twocolumn,aps,superscriptaddress,showpacs]{revtex4-2}

\usepackage{amssymb}
\usepackage{amsmath}
\usepackage{mathrsfs}
\usepackage{graphicx}
\usepackage{subfigure}
\usepackage[normalem]{ulem}
\usepackage[dvips]{color}

\begin{document}

\title{Effects of a first-order QCD phase transition on light nuclei production}
\author{He Liu}
\email{liuhe@qut.edu.cn}
\affiliation{Science School, Qingdao University of Technology, Qingdao 266520, China}
\author{Kai-Jia Sun}
\email{kjsun@fudan.edu.cn}
\affiliation{Key Laboratory of Nuclear Physics and Ion-beam Application (MOE), Institute of Modern Physics, Fudan University, Shanghai 200433, China}
\affiliation{Shanghai Research Center for Theoretical Nuclear Physics, NSFC and Fudan University, Shanghai 200438, China}
\author{Peng-Cheng Chu}
\email{kyois@126.com}
\affiliation{Science School, Qingdao University of Technology, Qingdao 266520, China}
\date{\today}
\begin{abstract}
Using an extended Polyakov-looped Nambu--Jona-Lasinio (PNJL) model to describe the baryon density fluctuations of quark matter along the isentropic trajectories corresponding to different $s/\rho_B$ values extracted from Au+Au collisions at energies $\sqrt{s_{NN}} = 7.7-200$ GeV, we investigate the effects of the first-order phase transition on the light nucleus yield ratio $N_t \times N_p/N_d^2$. The results indicate that the second-order scaled density moment $y_2$, used to quantify density fluctuations, rapidly increases to form a peak when the isentropic trajectories pass through the phase coexistence region. We extract the yield ratios $N_t\times N_p/N_d^2$ at chemical freeze-out from the isentropic trajectories at different collision energies and found significant enhancements at 19.6 GeV and 27 GeV. This is similar to the trends observed by the STAR experiment, suggesting that the enhancements in the yield ratios $N_t\times N_p/N_d^2$ observed in the STAR experiment could be explained by the density fluctuations generated in the first-order phase transition region. 
\end{abstract}

\maketitle
\textit{Introduction.} Exploring the phase diagram of Quantum Chromodynamics (QCD) is one of the main goals of relativistic heavy-ion collisions~\cite{Luo17,Bzd20}. Lattice QCD simulations indicate that the transition between the quark-gluon plasma (QGP) and the hadronic matter is a smooth crossover at nearly zero baryon chemical potential ($\mu_B\approx 0$)~\cite{Aok06,Gup11,Bor14}. Based on investigations from various effective models, such as the Nambu--Jona-Lasinio (NJL) model, as well as advanced functional methods including the Dyson-Schwinger equation (DSE) and the functional renormalization group (FRG)~\cite{Sch99,Zhu00,Fuk08,Che16,Liu16,Liu24,Qin11,Fu20}, the transition can be a first-order one at large $\mu_B$ region, resulting in a critical end point (CEP) on the first-order phase transition line. To search for the first-order phase transition and the potential CEP of QCD matter, experimental programs such as the Beam Energy Scan (BES-I and BES-II) have been carried out at the Relativistic Heavy-Ion Collider (RHIC) by varying the collision energy to cover a wide range of temperature $T$ and $\mu_B$ in the QCD phase diagram~\cite{Agg10,Ada142}. The STAR experiment has measured the energy dependence of observables that are sensitive to the CEP and/or first-order phase transition. Nonmonotonic energy dependencies were observed in these observables like net-proton fluctuations~\cite{Ada21L,Abd21}, pion HBT radii~\cite{Ada15,Ada21C}, baryon directed flow~\cite{Ada141,Ada18}, intermittency of charged hadrons~\cite{Abd23B}, as well as light nucleus yield ratios~\cite{Abd23L}.  

Light nucleus production in heavy-ion collisions is an active area of research both experimentally~\cite{Coc60,Arm00,Rei10,Ada16,Ada19} and theoretically~\cite{Cse86,Sun18,Sun21B,Zha20}. The yield ratios $N_t \times N_p/N_d^2$ of tritons ($N_t$), deuterons ($N_d$), and protons ($N_p$) at $\sqrt{s_{NN}} = 7.7-200$ GeV have been measured by the STAR experiment at RHIC. In $0-10\%$ most central Au + Au collisions at $\sqrt{s_{NN}} = 19.6$ and 27 GeV, the yield ratios $N_t \times N_p/N_d^2$ show significant enhancements relative to the coalescence baseline with a combined significance of 4.1$\sigma$~\cite{Abd23L}. Studies based on the coalescence models suggest that the fluctuations of neutron (baryon) density distributions and nucleon density-density correlations both lead to enhancements of the light nucleus yield ratios through the relation $N_t\times N_p/N_d^2\approx \frac{1}{2\sqrt{3}}[1+\Delta \rho_n +\frac{\lambda}{\sigma}G(\frac{\xi}{\sigma})]$, where $\Delta \rho_n$ is the average neutron density fluctuation, $\xi$ is the correlation length, and the term $\frac{\lambda}{\sigma}G(\frac{\xi}{\sigma})$ denotes the contribution from the long-range correlation between neutrons and proton, which increases monotonically with $\xi$~\cite{Sun18,Sun21B}. When the system approaches CEP, although the density fluctuation $\Delta \rho_n$ in a homogeneous system vanishes, the correlation length $\xi$ becomes divergent. However, during the first-order phase transition, the system could exhibit significant large density fluctuations~\cite{Ste12}, despite the correlation lengths within each phase being much smaller than those near the critical point. Hydrodynamic~\cite{Ste12,Ste13,Her13,Her14} and transport models~\cite{Li17,Sun22} also reveal that spinodal instability amplifies initial inhomogeneities in heavy-ion collisions as matter traverses the first-order transition region. Therefore, both first-order transitions and critical proximity can enhance the yield ratio $N_t \times N_p / N_d^2$, provided these effects persist through the hadronic stage of the collision. This raises a question: Are the enhancements of light nucleus yield ratios observed by the STAR experiment at 19.6 and 27 GeV related to the first-order phase transition or critical region?

\textit{Methods.} In this work, we focus on investigating the effect of a first-order phase transition on baryon density fluctuations in QCD phase diagram based on an extended Polyakov-looped Nambu--Jona-Lasinio (PNJL) model~\cite{Nam61A,Nam61B,Bha17,Li19} including three eight-quark interactions~\cite{Bha17,Li19,Bha10,Sun21D,Yan24}. To theoretically quantify the baryon density fluctuations, we consider using the second-order scaled density moment $y_2$ as the physical quantity for our study, which is defined as $y_2 = \bar{\rho_B^2}/\bar{\rho_B}^2 = [\int d\mathbf{x}\rho_B(\mathbf{x})][\int d\mathbf{x} \rho_B^3(\mathbf{x})] /[\int d\mathbf{x}\rho_B^2(\mathbf{x})]^2$, where $\rho_B(\mathbf{x})$ denotes the baryon density distribution in coordinate space. As shown in Refs.~\cite{Sun17,Sun18,Sun21B}, the quantity $y_2\approx 1+\Delta \rho_B$ is directly related to the light nucleus yield ratios in heavy-ion collisions. The yield ratio is approximately given by $y_2/(2\sqrt{3})$ when the correlation length is negligible. If the evolution of matter in the RHIC-BES can be approximated as an adiabatic expansion process, we can calculate the density moment $y_2$ for isentropic trajectories corresponding to values of $s/\rho_B$ within the ongoing RHIC-BES energy range and extract the yield ratios at different collision energies. In this way, we can theoretically reproduce the enhancements of yield ratios $N_t \times N_p/N_d^2$ in heavy-ion collisions at $\sqrt{s_{NN}}=19.6$ and $27$ GeV.  

\newcommand{\thickhline}{
\noalign{\hrule height 2\arrayrulewidth}
}
\begin{table}
\centering 
\caption{Parameters for the Polyakov-loop potential in the extended PNJL model}
\begin{tabular*}{0.48\textwidth}{@{\extracolsep{\fill}} *{7}{c}} 
\thickhline
$T_0$(MeV) & $a_0$ & $a_1$ & $a_2$ & $b_3$ & $b_4$ & $\kappa$ \\ 
\hline 
175 & 6.75 & -9.8 & 0.26 & 0.805 & 7.555 & 0.1\\ 
\thickhline
\end{tabular*}
\end{table}

The Lagrangian density of the realistic 3-flavor PNJL model, incorporating three eight-quark interactions, can be given by\cite{Bha17,Li19,Sun21D,Yan24} 
\begin{eqnarray}\label{eq1}
\mathcal{L}_{\textrm{PNJL}}^{\textrm{SU(3)}} &=& \bar{\psi}(i\gamma^{\mu}D_{\mu}-\hat{m})\psi 
+G_{S}\sum_{a=0}^{8}[(\bar{\psi}\lambda_{a}\psi)^{2}+(\bar{\psi}i\gamma_{5}\lambda_{a}\psi)^{2}] 
\notag\\
&-& K\{\det[\bar{\psi}(1+\gamma_{5})\psi]+\det[\bar{\psi}(1-\gamma_{5})\psi]\}
\notag\\
&+&\mathcal{L}_1^{8q}+\mathcal{L}_2^{8q}+\mathcal{L}_{SV}^{8q}+U^{'}(\Phi[A],\bar{\Phi}[A],T),
\end{eqnarray}
where $\psi = (u, d, s)^T$ represents the quark fields with three flavors, $\hat{m} = diag(m_u,m_d,m_s)$ is the current quark mass matrix, and $\lambda_a$ are the flavor SU(3) Gell-Mann matrices with $\lambda_0 = \sqrt{2/3}I$. The covariant derivative is defined by $D_{\mu}=\partial_{\mu}-iA_{\mu}$ with the background gluon field $A_{\mu}=\delta_{\mu,0}A_0$ supposed constant and uniform. The temperature-dependent Polyakov effective potential $U^{'}(\Phi[A],\bar{\Phi}[A],T)$ is a function of the Polyakov loop $\Phi[A]$ and its Hermitian conjugate $\bar{\Phi}[A]$ and it reads \cite{Bha17,Li19}
\begin{eqnarray}\label{eq2} 
\frac{U^{'}}{T^4}=\frac{U}{T^4}-\kappa \textrm{ln}[J(\Phi,\bar{\Phi})],
\end{eqnarray}
where
\begin{eqnarray}\label{eq3} 
\frac{U}{T^4}=-\frac{b_2(T)}{2}\Phi\bar{\Phi}-\frac{b_3}{6}(\Phi^3+\bar{\Phi}^3)+\frac{b_4}{4}(\Phi\bar{\Phi})^2,
\end{eqnarray}
and
\begin{eqnarray}\label{eq4} 
J=(\frac{27}{24\pi^2})(1-6\Phi\bar{\Phi}+4(\Phi^3+\bar{\Phi}^3)-3(\Phi\bar{\Phi})^2).
\end{eqnarray}
The coefficient $b_2(T)$ is chosen to have a temperature dependence of the form
\begin{eqnarray}\label{eq5}
b_2(T)=a_0+a_1 \frac{T_0}{T} \textrm{exp} (-a_2 \frac{T}{T_0}),
\end{eqnarray}
and $T_0$, $a_0$, $a_1$, $a_2$, $b_3$, $b_4$ and $\kappa$ are chosen to be constants and the details are listed in Table I.

\begin{table}
\centering 
\caption{Parameters in the extended PNJL model}
\begin{tabular*}{0.48\textwidth}{@{\extracolsep{\fill}} *{4}{c}} 
\thickhline
$m_{u,d}$(MeV) & $m_s$(MeV)& $\Lambda$(MeV) & $G_S\Lambda^2$ \\ 
\hline 
5.5 & 183.468 & 637.720 & 2.914 \\ 
\thickhline
$K\Lambda^5$ & $G_1$(MeV$^{-8}$)& $G_2$(MeV$^{-8}$)& $G_{SV}\Lambda^8$ \\ 
\hline 
9.496& $2.193\times 10^{-21}$&$-5.890\times 10^{-22}$&-3500.00 \\ 
\thickhline
\end{tabular*}
\end{table}

In above Eq. (1), $G_S$ is the scalar coupling constant, and the $K$ term represents the six-point Kobayashi-Maskawa-t' Hooft (KMT) interaction that breaks the axial symmetry $U(1)_A$~\cite{Hoo76}. $\mathcal{L}_1^{8q}$ and $\mathcal{L}_2^{8q}$ are the eight-quark interaction terms described by~\cite{Bha17,Li19}
\begin{eqnarray}\label{eq6}
\mathcal{L}_1^{8q}=\frac{G_1}{2}\{[\bar{\psi}_i(1+\gamma_5)\psi_j][\bar{\psi}_j(1-\gamma_5)\psi_i]\}^2,
\end{eqnarray}
and
\begin{eqnarray}\label{eq7}
\mathcal{L}_2^{8q}&=&G_2\{[\bar{\psi}_i(1+\gamma_5)\psi_j][\bar{\psi}_j(1-\gamma_5)\psi_k]
\notag\\
&\times&[\bar{\psi}_k(1+\gamma_5)\psi_l][\bar{\psi}_l(1-\gamma_5)\psi_i]\}.
\end{eqnarray}
These two the eight-quark interaction terms with coupling constants $G_1$ and $G_2$ in the Lagrangian could solve the problem of unstable vacuum in (P)NJL Model~\cite{Bha10,Bha17}. The main effect of the eight-quark interaction terms is to shift the crossover region to $T_0\approx 165$ MeV at vanishing baryon chemical potential, thus bringing them closer to lattice QCD results and the experimental freeze-out line. The term $\mathcal{L}_{SV}^{8q}$ is the third eight-quark scalar-vector interaction given by~\cite{Sun21D,Yan24}
\begin{eqnarray}\label{eq8}
\mathcal{L}_{SV}^{8q}&=&G_{SV}\{\sum_{a=1}^{3}[(\bar{\psi}\lambda^{a}\psi)^{2}+(\bar{\psi}i\gamma_{5}\lambda^{a}\psi)^{2}]\} 
\notag\\
&\times&\{\sum_{a=1}^{3}[(\bar{\psi}\gamma^{\mu}\lambda^{a}\psi)^{2}+(\bar{\psi}\gamma_{5}\gamma^{\mu }\lambda^{a}\psi)^{2}]\}.
\end{eqnarray}
Although the scalar-vector coupled interaction has no effects on the QCD vacuum properties and on the phase transition boundary in the QCD phase diagram, the strength of the interaction $G_{SV}$ can change the critical temperature and baryon chemical potential~\cite{Sun21D}. As shown in Refs.~\cite{Sun21D,Sun22,Yan24}, with a negative $G_{SV}$ the critical point moves to higher temperature and lower baryon chemical potential. Therefore, the range of spinodal region by varying the three eight-quark coupling constants ($G_1$, $G_2$, and $G_{SV}$) can cover sufficiently the region that can be probed in realistic heavy-ion collisions. In this work, the different parameters in the extended PNJL model are given in Table II~\cite{Bha17,Li19}. In the appendix, we discuss the influence of different parameters on the phase diagram as well as provide the rationale for parameter selection.
 
\begin{table}
\centering 
\caption{Estimate of the initial entropy density and of the entropy per baryon in Au+Au collisions at different collision energies. The kinetic and chemical freeze-out temperatures are obtained in Refs.~\cite{Ada17C,Abe09C,Mot20}.}
\begin{tabular*}{0.48\textwidth}{@{\extracolsep{\fill}} *{5}{c}} 
\thickhline 
$\sqrt{s_\textrm{NN}}$(GeV) & $s_0$(fm$^{-3}$)& $s/\rho_B$ & $T_{\textrm{kin}}^{\textrm{fo}}$ & $T_{\textrm{ch}}^{\textrm{fo}}$ \\ \hline 
7.7  & 29.6 & 17.5  & 116 &143 \\
11.5 & 35.3 & 26.7  & 118 &151 \\
19.6 & 43.0 & 45.8  & 113 &158 \\
27.0 & 45.8 & 56.8  & 117 &160 \\
39.0 & 47.6 & 84.3  & 117 &160 \\
62.4 & 60.2 & 123.9 & 99  &161 \\
200  & 84.0 & 331.6 & 89  &162 \\
\thickhline 
\end{tabular*}
\end{table}

\textit{Density fluctuations and enhancements of light nucleus yield ratios.} We first display the phase diagram from the PNJL model for quark matter in $T-\rho_B$ plane in Fig. 1. The shaded region respresents the mechanical (spinodal) instability region with $(\partial P /\partial \rho_B)_T < 0$. As a feature of spinodal instability, infinitesimal density fluctuations within this unstable region spontaneously drive phase separation~\cite{Mul95}. The gray dash-dotted line depicts the boundary of the two-phase coexistence region. For the liquid-gas-like phase transition, e.g., the (P)NJL model, the chemical potential in the liquid-gas coexistence phase can be regarded as that of the first-order chiral phase transition boundary in $T-\mu_B$ plane. The black dashed line and dot are respectively for the chiral crossover and CEP that connect the chiral crossover and coexistence region. It can be observed that due to the effect of the coupling constants $G_1$ and $G_2$, the crossover temperature at vanishing baryon density is approximately 165 MeV, which is consistent with the predictions from Lattice QCD calculations and experimental results~\cite{Din15,Abe09C,Baz14,Ada17C}. More importantly, with the vector-scalar interaction coupling constant $G_{SV}=-3500\Lambda^{-8}$, the CEP temperature ($T_c> 150$ MeV) is significantly increased compared to the default PNJL model, resulting in a larger coexistence region, allowing the evolution trajectory of higher-energy heavy-ion collisions to pass through the coexistence region. In heavy-ion collisions, the produced system undergoes an approximately isentropic expansion driven by pressure gradients along trajectories of constant entropy per baryon ($s/\rho_B$=const), when dissipative effects such as viscosity, heat conduction, and charge diffusion are neglected. In order to provide phenomenological analysis for the BES program at RHIC, we need to estimate the entropy per baryon and the initial entropy density of the system at various Au+Au collision energies from 7.7 to 200 GeV. The initial entropy density $s_0$ could be estimated by assuming that its value is proportional to the measured rapidity density of charged particles $dN_{ch}/d\eta$~\cite{Mot20}. For Au+Au collisions at $\sqrt{s_{NN}} = 200$ GeV the value of $s_0$ is about 84 fm$^{-3}$~\cite{Alb11}, and the initial entropy density at lower collision energies is obtained by rescaling the estimate at $\sqrt{s_{NN}} = 200$ GeV according to $dN_{ch}/dy$~\cite{Ada17C,Abe09C}. If using $s_0$ as the initial condition in hydrodynamic calculations can successfully reproduce the soft hadron distributions, then the $s/\rho_B$ ratios at different collision energies can be extracted from the yields of identified hadrons ($\pi^{\pm}, K^{\pm}, p/\bar{p}$). As shown in Ref.~\cite{Mot20}, the results for $s_0$ and $s/\rho_B$ at $\sqrt{s_{NN}} = 7.7-200$ GeV are collected in Table III, where we also display the values of the kinetic and chemical freeze-out temperatures obtained in Ref.~\cite{Ada17C,Abe09C}. In Fig. 1, we depict the isentropic trajectories of the $s/\rho_B$ values listed in Table II using solid lines of different colors. In addition, the isentropic lines with $s/\rho_B = 5, 10$ are shown for comparison. It can be seen that, with the coupling constant set in our model, the trajectories corresponding to $s/\rho_B < 56.8$ for collision energies below 39 GeV pass through the coexistence region of the phase diagram, which could lead to enhanced density fluctuations in the systems produced at these collision energies. 

 \begin{figure}[tbh]
\includegraphics[scale=0.38]{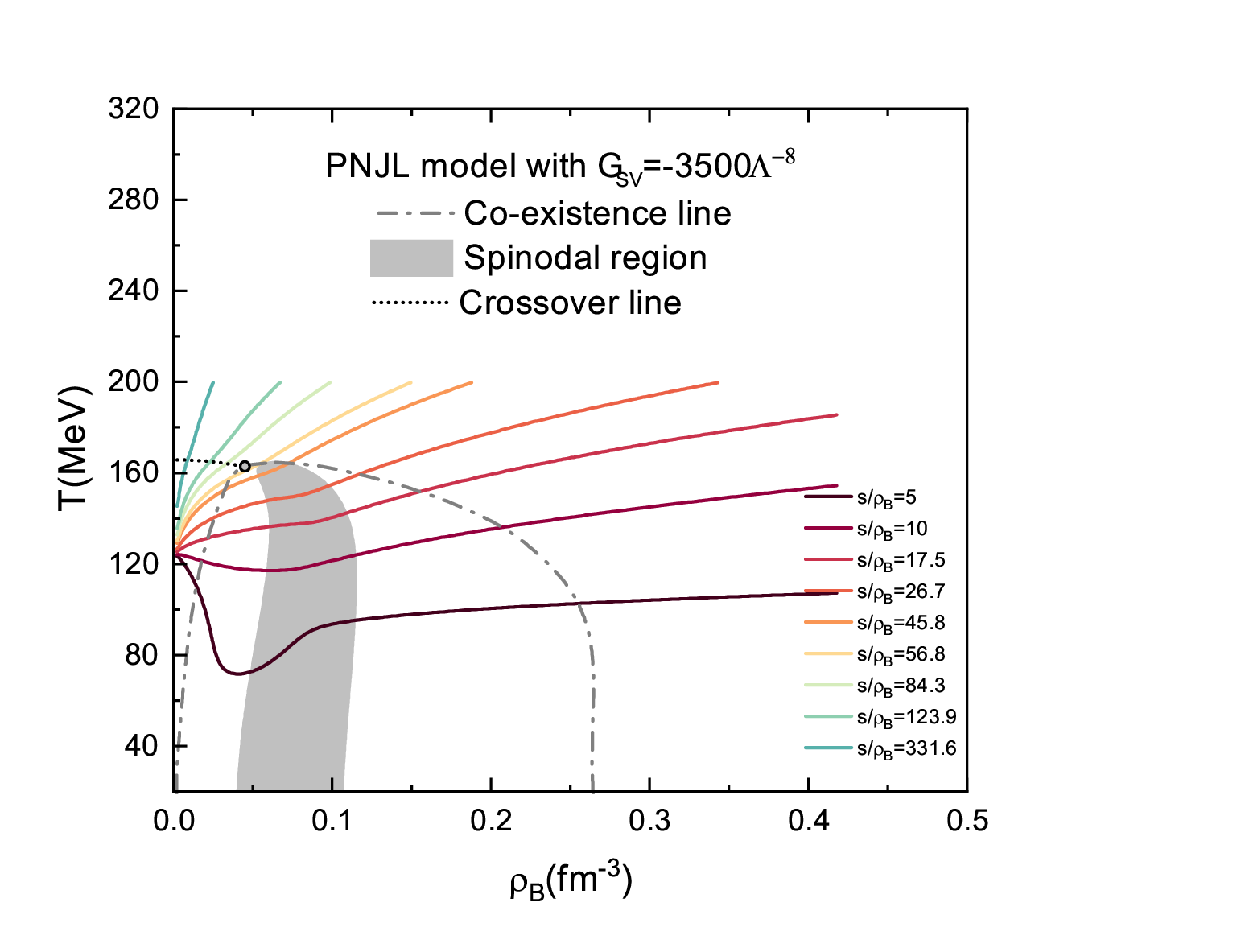}
\caption{Phase diagram from an extended PNJL model for quark matter in $T-\rho_B$ plane. The gray dash-dotted line represents the boundary of the two-phase coexistence region, and the shaded region is the mechanical (spinodal) instability regions with $(\partial P /\partial \rho_B)_T < 0$. The black dashed line and dot are respectively for the chiral crossover and CEP that connect the chiral crossover and coexistence region. The different color solid lines represent isentropic trajectories of the $s/\rho_B$ values listed in Table II.} \label{fig1}
\end{figure}

Spinodal instability can complicate the observation of first-order phase transitions in heavy-ion collisions. Dynamical simulations reveal that when the bulk of an evolving system traverses the spinodal region during a first-order transition, spinodal instabilities universally trigger phase separation within the coexistence region~\cite{Ran10,Li16,Li17}. This process amplifies density fluctuations, leading to clustering phenomena that enhance the yields of composite light nuclei, such as deuterons and tritons~\cite{Ste12,Sun21A}. As demonstrated in Refs.~\cite{Sun18,Sun21B} using the nucleon coalescence model, the yield ratio $N_t \times N_p / N_d^2$ is approximately given by the scaled density moment $y_2 / (2\sqrt{3})$ under the condition of zero density-density correlation length. For a system undergoing phase separation into two distinct phases, the second-order scaled density moment is expressed as $y_2 = (\rho_1 \lambda_1 + \rho_2 \lambda_2)(\rho_1^3 \lambda_1 + \rho_2^3 \lambda_2) / (\rho_1^2 \lambda_1 + \rho_2^2 \lambda_2)^2$, where $\rho_i$ and $\lambda_i$ ($i=1,2$) denote the baryon densities and volume fractions of the dense and dilute phases, respectively~\cite{Li17}. Within the coexistence region, the equilibrium densities $\rho_1$ and $\rho_2$ are determined by the Maxwell construction. It is worth noting that the hot medium created in the heavy-ion collisions (HICs) is a dynamically expanding system. Especially when the medium passes through a first-order phase transition, there may exist highly nontrivial processes such as supercooling or bubble formations~\cite{Ste12,Ste13}. As stated in Ref.~\cite{Mis99} that the nonequilibrium phase transitions in rapidly expanding matter during relativistic HICs may lead to the fragmentation of the plasma phase into droplets surrounded by undersaturated hadronic gas. Describing these nonequilibrium phase transition features by thermodynamic methods proves highly challenging. Accordingly, our calculation of density moment $y_2$ employs a simplifying approach grounded in the standard thermodynamical concepts. 

\begin{figure}[tbh]
\centering
\includegraphics[scale=0.33]{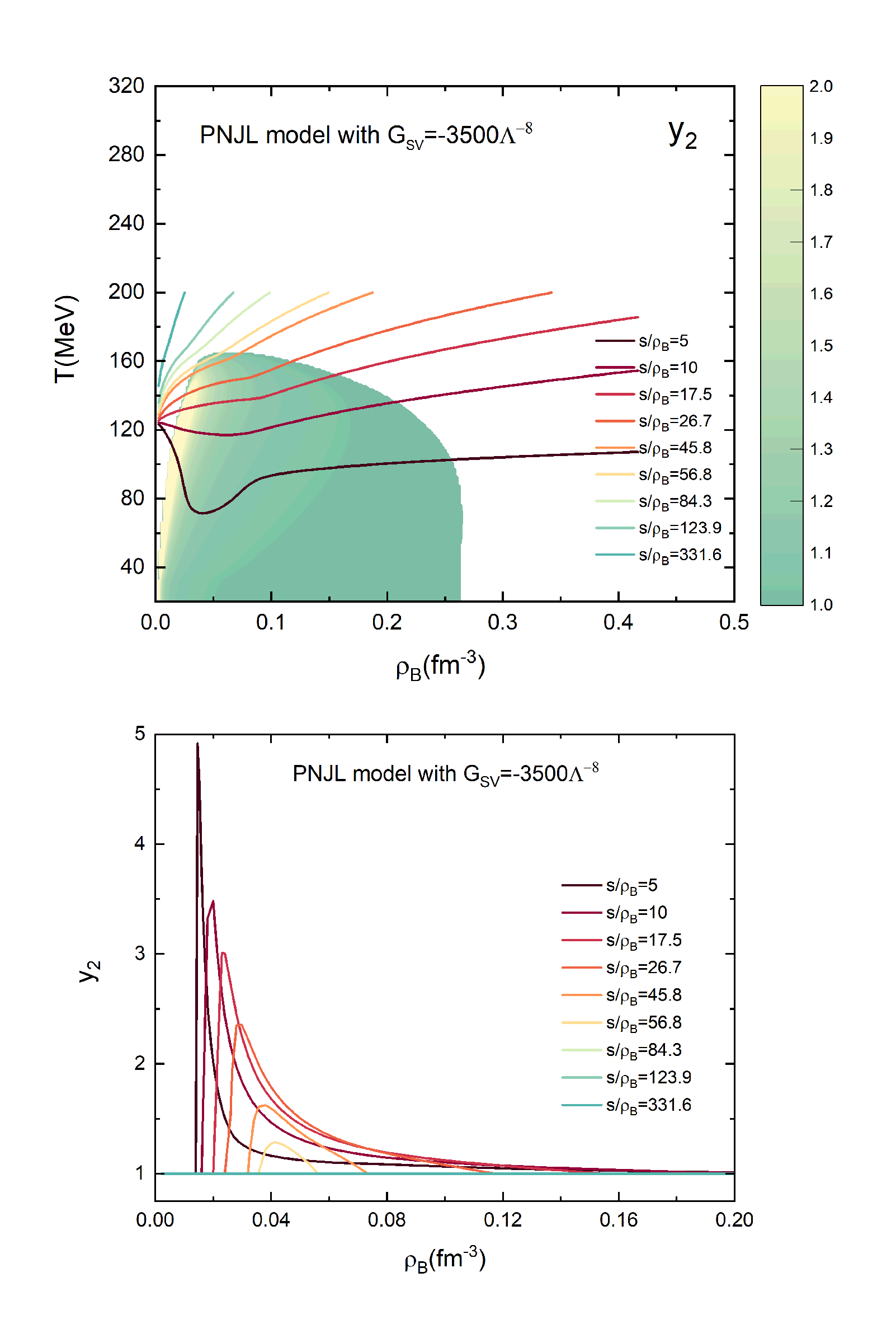}
\caption{Upper panel: the contour map of $y_2$ in the full phase diagram based on the extended PNJL model with the coupling constant \small{$G_{SV}=-3500\Lambda^{-8}$}. The different color solid lines represent isentropic trajectories of the $s/\rho_B$ values for different collision energies listed in Table II. Besides, the isentropic lines $s/\rho_B$ = 5, 10 are shown for comparison. Lower panel: $y_2$ as functions of $\rho_B$ along the isentropic trajectories of the $s/\rho_B$ values for different collision energies listed in Table II.} \label{fig2}
\end{figure}

In the upper panel of Fig. 2, we present the contour map of $y_2$ in the full phase diagram based on the extended PNJL model with the coupling constant $G_{SV}=-3500\Lambda^{-8}$. It can be seen that $y_2$ clearly delineates the extent of the coexistence region. In the non-coexistence (white) region, $y_2$ takes the constant value $y_2=1$ due to the absence of density fluctuations, whereas $y_2$ is significantly enhanced in the coexistence (colored) region, especially on the left side of the coexistence region. To better describe the changes in density fluctuations at different energies during the evolution of heavy-ion collisions, the results for $y_2$ as functions of $\rho_B$ along the isentropic trajectories of the $s/\rho_B$ values are shown in the lower panel of Fig. 2 by the different colored solid lines. It can be seen that as the isentropic trajectories pass through the coexistence region, $y_2$ rapidly increases to form a peak and then gradually decreases back to 1. We can also observe that the peak value of $y_2$ increases as the entropy per baryon (collision energy) decreases. The increase in density fluctuations leads to enhanced clustering, which facilitates the production of light nuclei. Therefore, the yield of light nuclei increases as the collision energy decreases. The experiments that the energy dependence of $N_d/N_p$ and $N_t/N_p$ ratios in the midrapidity of central heavy-ion collisions from the FOPI~\cite{Rei07}, E864~\cite{Arm00}, PHENIX~\cite{Adl05}, and ALICE~\cite{Ada16} also confirm this point. The results indicate both $N_d/N_p$ and $N_t/N_p$ ratios increase monotonically with decreasing collision energy. In addition, the coalescence model, which predicts light nucleus production at midrapidity based on baryon density ($\rho_B$) via the relationship $N_A/N_p \propto \rho_B^{A-1}$, can also describe similar energy dependence trends~\cite{Zha21}

\begin{figure}[tbh]
\includegraphics[scale=0.34]{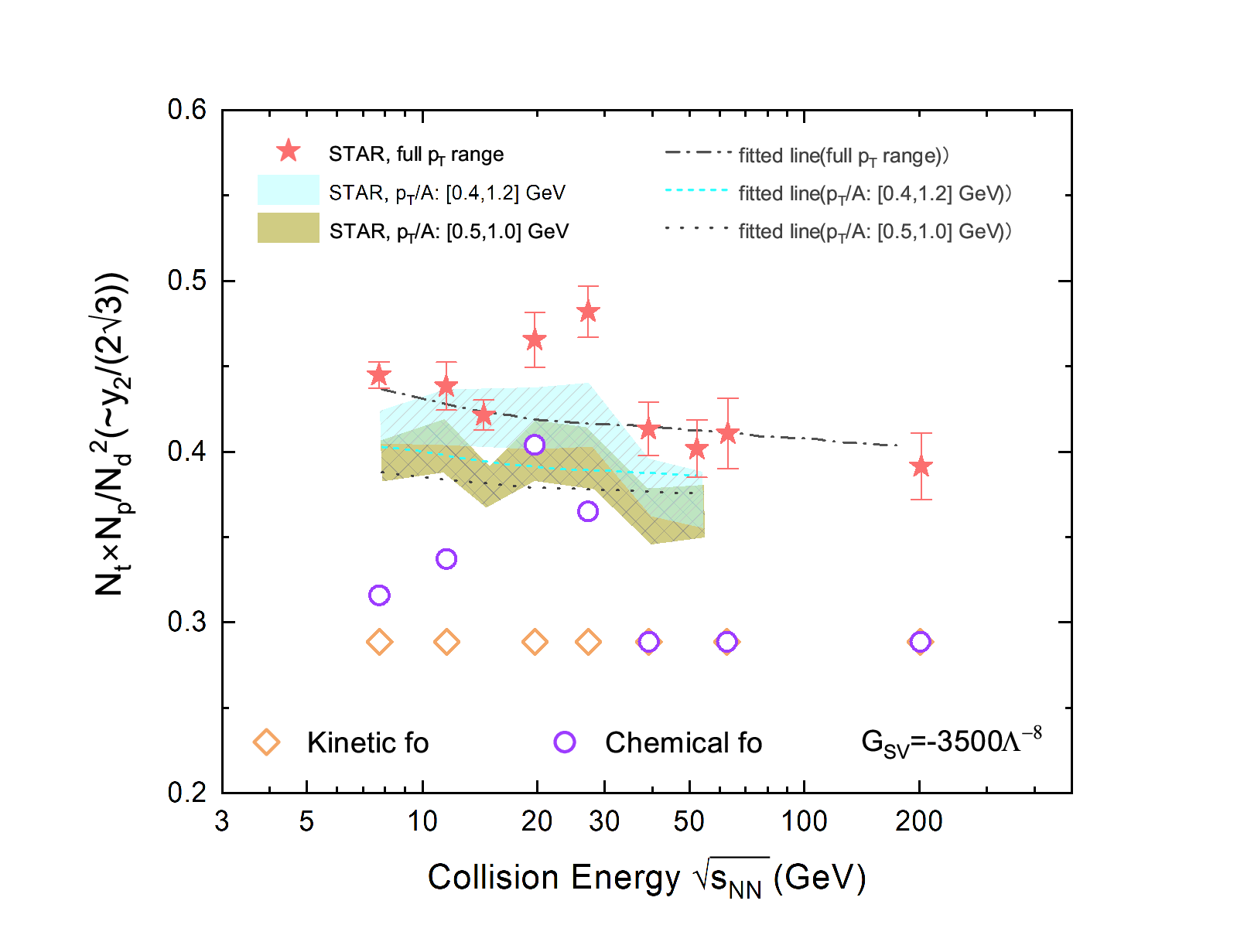}
\caption{Collision energy dependence of the yield ratio $N_t\times N_p/N_d^2$, where the yield ratio is approximated by $y_2/(2\sqrt{3})$. The violet open circles correspond to the results at chemical freeze-out during the isentropic process at different energies, while the orange open squares represent the results at kinetic freeze-out. The results from 0\%-10\% central Au + Au collisions at RHIC are also shown for comparison. Red solid stars with error bars are the final results with extrapolation to the full $p_T$ range and the dash-dotted fitted lines are the coalescence baselines from the coalescence-inspired fit. The colored bands and dashed fitted lines denote $p_T$ acceptance dependence, for which the statistical and systematic uncertainties are added in quadrature.} \label{fig3}
\end{figure}

As mentioned in \textit{Introduction}, the yield ratio $N_t\times N_p/N_d^2$ is predicted to be sensitive to the local baryon density fluctuations and critical correlation lengths, which could be used to probe the first-order phase transition and/or CEP in heavy-ion collisions. To determine whether the enhancements of light nucleus yield ratios observed by the STAR experiment at 19.6 and 27 GeV is related to a first-order phase transition, we present in Fig. 3 the collision energy dependence of the yield ratio $N_t\times N_p/N_d^2$, where the yield ratio is approximated by $y_2/(2\sqrt{3})$, neglecting the effect of the correlation length. The violet open circles correspond to the results at chemical freeze-out during the isentropic process at different energies, while the orange open squares represent the results at kinetic freeze-out. The temperatures for chemical and kinetic freeze-out are obtained from Ref.~\cite{Ada17C,Abe09C}, as shown in Table II. The results from 0-10\% central Au + Au collisions at RHIC are also shown for comparison. Our results indicate that the yield ratios $N_t\times N_p/N_d^2$ at kinetic freeze-out have no significant enhancements for different collision energies. This occurs because the kinetic freeze-out temperature is lower than that of all isentropic trajectories with $s/\rho_B >19.5$, which indicates that the QCD phase transition has already concluded by the time of kinetic freeze-out. Encouragingly, the enhancements of yield ratios at chemical freeze-out occur at 19.6 and 27 GeV, which is similar to the trends observed by the STAR experimental. This suggests that the enhancements in the yield ratios $N_t\times N_p/N_d^2$ observed in the STAR experiment can be attributed to the density fluctuations generated in the first-order phase transition region. However, we also note that the overall ratios obtained from our calculations are lower than the experimental results and the baseline provided by the coalescence model, which is due to the neglect of the density-density correlation length. Recently, the non-monotonic dependence of net-proton fluctuations, which are sensitive to the correlation length, on collision energy has indeed been observed in the preliminary data from the STAR Collaboration, particularly the dip structure around 19.6 GeV~\cite{Abo25}. Therefore, we cannot rule out the possibility that the critical region could also lead to the enhancements in the yield ratios in the range of 19.6 to 27 GeV. In future work, we will further explore the effects of the correlation length in the critical region on the light nucleus yields ratios, and integrate the extended PNJL model and its parameters into transport models to investigate the structure of the QCD phase diagram. 

\textit{Summary and outlook.} Based on an extended Polyakov-looped Nambu--Jona-Lasinio (PNJL) model with three eight-quark interactions, we investigate the effects of the two-phase coexistence region during a first-order phase transition on light nucleus production in QCD phase diagram. The range of coexistence region by varying the three eight-quark coupling constants ($G_1$, $G_2$, and $G_{SV}$) can cover sufficiently the region that can be probed in realistic heavy-ion collisions. Considering the evolution of matter in heavy-ion collisions as an isentropic process, we use the second-order scaled density moment $y_2$ approximation to calculate the baryon density fluctuations along the isentropic trajectories corresponding to different $s/\rho_B$ values extracted from Au+Au collisions at energies ranging from 7.7 to 200 GeV at RHIC. Results indicate that $y_2$ rapidly increases to form a peak when the isentropic trajectories pass through the coexistence region due to the density fluctuations in the first-order phase transition region. We extracted the yield ratios $N_t\times N_p/N_d^2$ at chemical freeze-out from the isentropic trajectories at different collision energies and found the significant enhancements at 19.6 GeV and 27 GeV, which is similar to the trends observed by the STAR experimental. This suggests that the enhancements in the yield ratios $N_t\times N_p/N_d^2$ observed in the STAR experiment could be attributed to the density fluctuations generated in the first-order phase transition region. Due to the lack of studies on the correlation length in the critical region, we cannot rule out the possibility that the critical region could also lead to the enhancements in the yield ratios in the range of 19.6 to 27 GeV. Further studies from transport modeling of heavy-ion collisions will be conducted to confirm if the enhancements are due to large density fluctuations during a first-order phase transition. 

\textit{Acknowledgments.} This work is supported by the National Natural Science Foundation of China under Grants No. 12205158, and No. 11975132, as well as the Shandong Provincial Natural Science Foundation, China Grants No. ZR2021QA037, No. ZR2022JQ04, and No. ZR2019YQ01.

\clearpage
\begin{widetext}

\section*{Appendix}
This appendix reviews the thermodynamic potential and related quantities in the PNJL model, with a discussion on how different vector-scalar couplings affect the phase diagram. In the mean-field approximation, the thermodynamic potential for the realistic 3-flavor PNJL model is given by
\begin{eqnarray}\label{eq9}
\Omega_{\textrm{PNJL}} &=&-2N_c\sum_i\int_0^\Lambda\frac{d^3p}{(2\pi)^3}E_i-2T\sum_i\int\frac{d^3p}{(2\pi)^3}\{\ln[1+3\Phi e^{-\beta(E_i-\tilde{\mu}_i)}+3\bar{\Phi}e^{-2\beta(E_i-\tilde{\mu}_i)}+e^{-3\beta(E_i-\tilde{\mu}_i)}]
\notag\\
&+& \ln[1+3\bar{\Phi}e^{-\beta(E_i+\tilde{\mu}_i)}+3\Phi e^{-2\beta(E_i+\tilde{\mu}_i)}+e^{-3\beta(E_i+\tilde{\mu}_i)}]\}+G_S\sum_i\sigma_i^2-4K\sigma_u\sigma_d\sigma_s
\notag\\
&+& 3\frac{G_1}{2}(\sum_i\sigma_i^2)^2+3G_2\sum_i\sigma_i^4+3G_{SV}(\sigma_{u}+\sigma_{d})^{2}(\rho_{u}+\rho_{d})^{2}+U^{\prime}(\Phi,\bar{\Phi},T),
\end{eqnarray}
where the factor $2N_c=6$ represents the spin and color degeneracy of the quark, and $\beta = 1/T$ is the inverse of the temperature. $E_{i} =\sqrt{M_{i}^{2}+p^{2}}$ is the single-quark energy for three different flavor quarks, where $M_i$ is the constituent quark mass given by the gap equations
\begin{eqnarray}\label{eq10}
M_u&=&m_u-2G_S\sigma_u+2K\sigma_d\sigma_s-2G_1\sigma_u(\sum_{i}\sigma_i^2)-4G_2\sigma_u^3-2G_{SV}(\rho _{u}+\rho_{d})^{2}(\sigma_{u}+\sigma_{d}),
\notag\\
M_d&=&m_d-2G_S\sigma_d+2K\sigma_s\sigma_u-2G_1\sigma_d(\sum_{i}\sigma_i^2)-4G_2\sigma_d^3-2G_{SV}(\rho_{u}+\rho_{d})^{2}(\sigma_{u}+\sigma_{d}),
\notag\\
M_s&=&m_s-2G_S\sigma_s+2K\sigma_u\sigma_d-2G_1\sigma_s(\sum_{i}\sigma_i^2)-4G_2\sigma_s^3.
\end{eqnarray}
The effective chemical potential $\tilde{\mu}_i$ can be expressed as
\begin{eqnarray}\label{eq11}
\tilde{\mu}_{u}&=&\mu_{u}+2G_{SV}(\rho_{u}+\rho_{d})(\sigma_{u}+\sigma_{d})^{2},
\notag\\
\tilde{\mu}_{d}&=&\mu_{d}+2G_{SV}(\rho_{u}+\rho_{d})(\sigma_{u}+\sigma_{d})^{2},
\notag\\
\tilde{\mu}_{s}&=&\mu_{s}.
\end{eqnarray}
In our calculation, we set $\mu_u=\mu_d=\mu_s=\mu_B/3$. $\sigma_i=\langle \bar{\psi_i} \psi_i \rangle$ stands for quark condensate, which serves as the order parameter for the chiral phase transition. From the thermodynamic potential given by Eq. (9), the physical values for the condensates $\sigma_u$, $\sigma_d$, and $\sigma_s$ as well as Polyakov loop $\Phi$ and $\bar{\Phi}$ can be calculated considering the following equation 
\begin{eqnarray}\label{eq12}
\frac{\partial\Omega_{\textrm{PNJL}}}{\partial\sigma_u}=\frac{\partial\Omega_{\textrm{PNJL}}}{\partial\sigma_d}=\frac{\partial\Omega_{\textrm{PNJL}}}{\partial\sigma_s} =\frac{\partial\Omega_{\textrm{PNJL}}}{\partial\Phi}=\frac{\partial\Omega_{\textrm{PNJL}}}{\partial\bar{\Phi}}=0.
\end{eqnarray} 
The pressure, number density, and entropy density can be derived using the thermodynamic relations in the grand canonical ensemble as 
\begin{eqnarray}\label{eq13}
P=-\Omega_{\textrm{PNJL}}, \quad \rho_i=-\frac{\partial\Omega_{\textrm{PNJL}}}{\partial\mu_i}, \quad s=-\frac{\partial\Omega_{\textrm{PNJL}}}{\partial T}.
\end{eqnarray}
For a system of quarks, the baryon number density is given by $\rho_B=(\rho_u+\rho_d+\rho_s)/3$. 

\begin{figure}[tbh]
\includegraphics[scale=0.33]{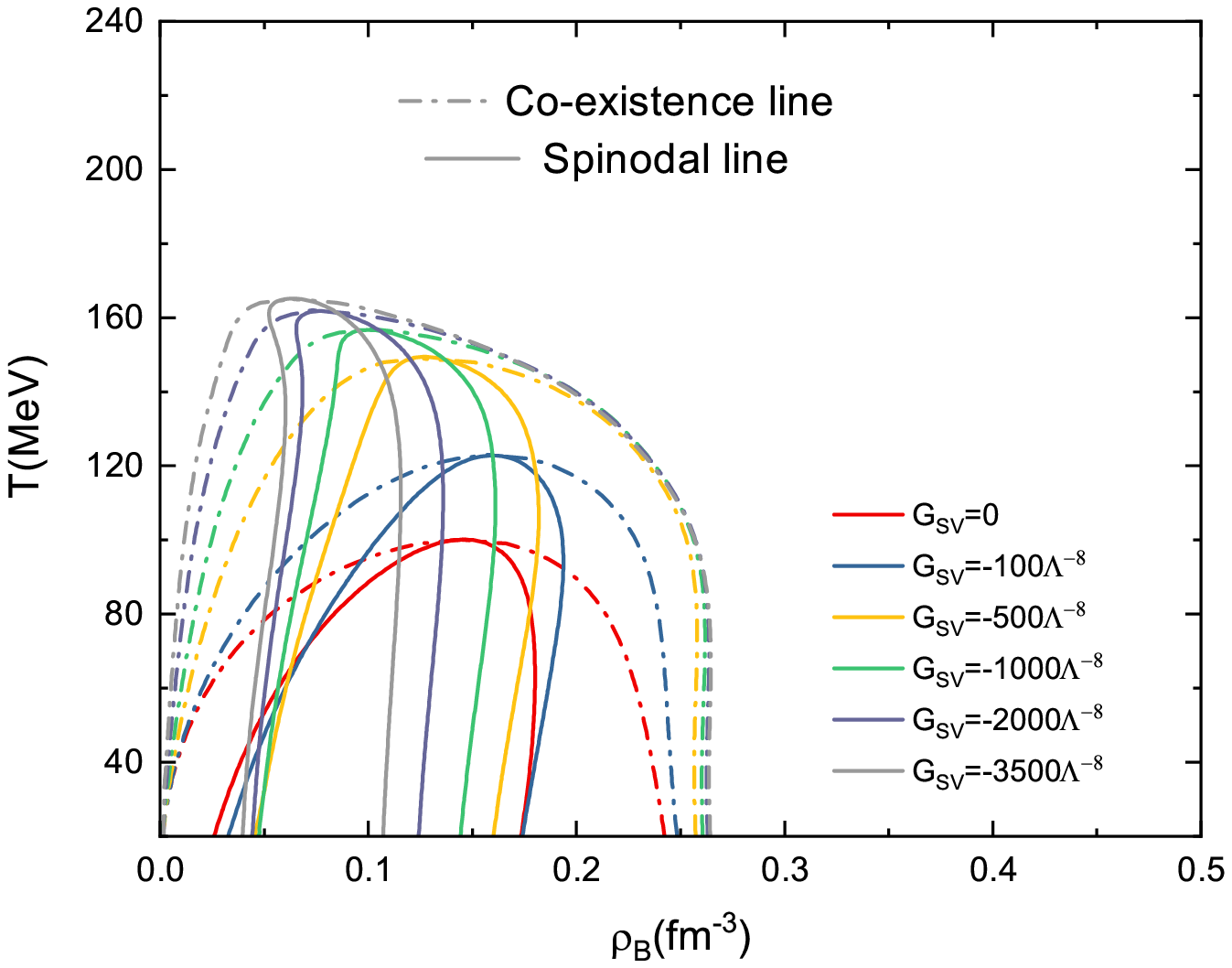}
\caption{Coexistence (dash-dotted) and spinodal (solid) lines in $T-\rho_B$ plane from the realistic 3-flavor PNJL model for different values of the scalar-vector coupling constant $G_{SV}$. Other parameters are set to the values in Table II.} \label{fig4}
\end{figure}
\begin{figure}[tbh]
\includegraphics[scale=0.33]{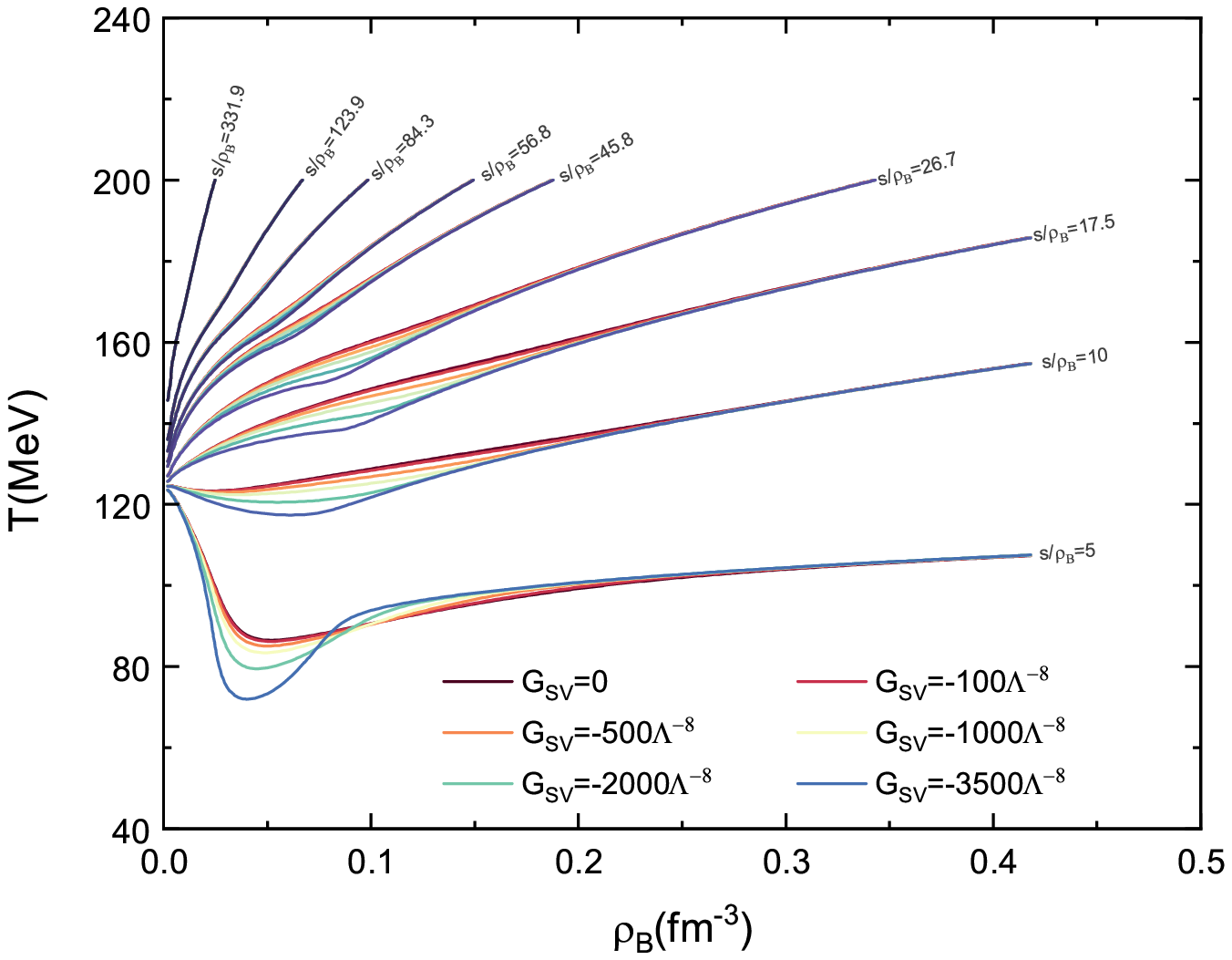}
\caption{Isentropic trajectories of the entropy per baryon ($s/\rho_B$) for the different collision energies in Table III, calculated for different values of the scalar-vector coupling constant $G_{SV}$.} \label{fig5}
\end{figure}

As mentioned earlier in the main text, we introduce three eight-quark interaction terms in the PNJL model to adjust the spinodal region in phase diagram so that it effectively covers the region probed by heavy-ion collision experiments. The two eight-quark interactions with coupling constants $G_1$ and $G_2$ are used to restrict the crossover region below $ T_0 \approx 165\ \mathrm{MeV}$, which is consistent with current lattice QCD calculations at zero baryon density and experimental predictions for the chemical freeze-out. We therefore adopt the parameter set from Refs.~\cite{Bha17,Li19} and keep the values of $G_1$ and $G_2$ fixed as summarized in Table II. 

Although the vector-scalar type eight-quark interaction does not influence the QCD vacuum properties and the phase transition boundary in the QCD phase diagram, its coupling strength $G_{SV}$ can significantly shift the position of the QCD critical point and modify the extent of the spinodal region. As shown in Fig. 4, the boundaries of the spinodal and coexistence regions vary with the coupling constant $G_{SV}$. A decrease in $G_{SV}$ broadens the coexistence region, shifts the spinodal region toward lower density, and enhances the CEP temperature. The effects of $G_{SV}$ can be understood using Eqs. (9)-(11). According to Eq. (9) and (11), the negative $G_{SV}$ resembles a vector interaction in the (P)NJL model, inducing a repulsive interaction among the quarks and stiffening the equation of state. Moreover, Eq. (11) shows that a negative $G_{SV}$ raises the chemical potential. Since the $G_{SV}$ term is density-dependent, it enlarges the coexistence region at finite density. On the other hand, Eq. (10) reveals that a negative $G_{SV}$ also counteracts the scalar coupling $G_S$. Being proportional to the square of the density, the $G_{SV}$ term significantly reduces the constituent quark mass in the finite density region. This effect drives the spinodal region toward lower densities and enhances the critical temperature. We further observe that as $G_{SV}$ is reduced below –500$\Lambda^{-8}$, its effect on the phase diagram begins to saturate. This is attributed to the dependence of the $G_{SV}$ term on both density and the quark condensate. Its impact reduces progressively as the spinodal region and critical point move to lower density regimes. Meanwhile, at very high densities, where the chiral symmetry is largely restored and the quark condensates approach zero, the effects of the $G_{SV}$ term become less important. This differs from the usual vector interaction in the (P)NJL model that becomes stronger at high densities.

In Fig. 5, we present the isentropic trajectories of the entropy per baryon ($s/\rho_B$) for the different collision energies in Table III. As can be seen from the figure, a negative $G_{SV}$ value softens the isentropic lines and the effect is more significant at lower collision energies (smaller $s/\rho_B$). A main goal of this work is to explain the enhanced yield ratios of light nuclei observed at 19.6 and 27 GeV, corresponding to $s/\rho_B = 45.8$ and 56.8. To achieve this, we reduced $G_{SV}$ to –3500$\Lambda^{-8}$, ensuring that the isentropic lines at $s/\rho_B = 45.8$ and 56.8 pass through the two phase co-existence region (as shown in Fig. 1), thereby inducing sufficiently strong density fluctuations in order to reproduce the experimental results. This mechanism provides the rationale for the parameter selection in Table II, including $G_1$, $G_2$, and $G_{SV}$.

\end{widetext}


\begin{thebibliography}{99}

\bibitem{Luo17} X. Luo and N. Xu, Nucl. Sci. Tech. 
\textbf{28}, 112 (2017).

\bibitem{Bzd20} A. Bzdak, S. Esumi, V. Koch, J. Liao, M. Stephanov, and N. Xu, Phys. Rept. 
\textbf{853}, 1 (2020).

\bibitem{Aok06} Y. Aoki, G. Endrődi, Z. Fodor, S.D. Katz, and K.K. Szabó, Nature (London) 
\textbf{443}, 675 (2006). 

\bibitem{Gup11} S. Gupta, X.F. Luo, B. Mohanty, H.G. Ritter, and N. Xu, Science 
\textbf{332}, 1525 (2011).

\bibitem{Bor14} S. Borsányi, Z. Fodor, C. Hoelbling, S.D. Katz, S. Krieg, and K.K. Sabzó, Phys. Lett. B 
\textbf{730}, 99 (2014).

\bibitem{Sch99} T.M. Schwarz, S.P. Klevansky, and G. Papp, Phys. Rev. C 
\textbf{60}, 055205 (1999).

\bibitem{Zhu00} P. Zhuang, M. Huang, and Z. Yang, Phys. Rev. C 
\textbf{62}, 054901 (2000).

\bibitem{Fuk08} K. Fukushima, Phys. Rev. D 
\textbf{77}, 114028 (2008);
 K. Fukushima, Phys. Rev. D 
\textbf{78}, 039902 (E) (2008).

\bibitem{Che16} J.W. Chen, J. Deng, H. Kohyama, and L. Labun, Phys. Rev. D 
\textbf{93}, 034037 (2016).

\bibitem{Liu16} H. Liu, J. Xu, L.W. Chen, K.J. Sun, Phys. Rev. D 
\textbf{94}, 065032 (2016).

\bibitem{Liu24} H. Liu, Y.H. Yang, C. Yuan, M. Ju, X.H. Wu, and P.C, Chu, Phys. Rev. D 
\textbf{109}, 074037 (2024).

\bibitem{Qin11} S.X. Qin, L. Chang, H. Chen, Y.X. Liu, and C.D. Roberts, Phys. Rev. Lett. 
\textbf{106}, 172301 (2011).

\bibitem{Fu20} W.J. Fu, J.M. Pawlowski, and F. Rennecke, Phys. Rev. D 
\textbf{101}, 054032 (2020).

\bibitem{Agg10} M. M. Aggarwal \textit{et al}., (STAR Collaboration), Phys. Rev. Lett. 
\textbf{105}, 022302 (2010).

\bibitem{Ada142} L. Adamczyk \textit{et al}., (STAR Collaboration), Phys. Rev. Lett. 
\textbf{112}, 032302 (2014). 

\bibitem{Ada21L} J. Adam \textit{et al}., (STAR Collaboration), Phys. Rev. Lett. 
\textbf{126}, 092301 (2021).
 
\bibitem{Abd21} M. Abdallah \textit{et al}., (STAR Collaboration), Phys. Rev. C 
\textbf{104}, 024902 (2021).

\bibitem{Ada15} L. Adamczyk \textit{et al}., (STAR Collaboration), Phys. Rev. C 
\textbf{92}, 014904 (2015). 

\bibitem{Ada21C} J. Adam \textit{et al}., (STAR Collaboration), Phys. Rev. C 
\textbf{103}, 034908 (2021).

\bibitem{Ada141} L. Adamczyk \textit{et al}., (STAR Collaboration), Phys. Rev. Lett. 
\textbf{112}, 162301 (2014). 

\bibitem{Ada18} L. Adamczyk \textit{et al}., (STAR Collaboration), Phys. Rev. Lett. 
\textbf{120}, 062301 (2018).

\bibitem{Abd23B} M.I. Abdulhamid \textit{et al}., (STAR Collaboration), Phys. Lett. B 
\textbf{845}, 138165 (2023).

\bibitem{Abd23L} M.I. Abdulhamid \textit{et al}., (STAR Collaboration), Phys. Rev. Lett. 
\textbf{130}, 202301 (2023). 

\bibitem{Coc60} V.T. Cocconi, T. Fazzini, G. Fidecaro, M. Legros, N.H. Lipman, and A.W. Merrison, Phys. Rev. Lett. 
\textbf{5}, 19 (1960).

\bibitem{Arm00} T.A. Armstrong, K.N. Barish, S. Batsouli, S.J. Bennett, M. Bertaina \textit{et al}., (E864 Collaboration), Phys. Rev. C 
\textbf{61}, 064908 (2000).

\bibitem{Rei10} W. Reisdorf \textit{et al}., (FOPI Collaboration), Nucl. Phys. A
\textbf{848}, 366 (2010).

\bibitem{Ada16} J. Adam \textit{et al}., (ALICE Collaboration), Phys. Rev. C 
\textbf{93}, 024917 (2016).

\bibitem{Ada19} J. Adam \textit{et al}., (STAR Collaboration), Phys. Rev. C 
\textbf{99}, 064905 (2019).

\bibitem{Cse86} L.P. Csernai and J.I. Kapusta, Phys. Rep. 
\textbf{131}, 223 (1986).

\bibitem{Sun18} K.J. Sun, L.W. Chen, C.M. Ko, J. Pu, and Z. Xu, Phys. Lett. B 
\textbf{781}, 499 (2018).

\bibitem{Sun21B} K.J. Sun, F. Li, and C.M. Ko, Phys. Lett. B 
\textbf{816}, 136258 (2021).

\bibitem{Zha20} W. Zhao, C. Shen, C.M. Ko, Q. Liu, and H. Song, Phys. Rev. C 
\textbf{102}, 044912 (2020).

\bibitem{Ste12} J. Steinheimer and J. Randrup, Phys. Rev. Lett. 
\textbf{109}, 212301 (2012)


\bibitem{Ste13} J. Steinheimer and J. Randrup, Phys. Rev. C 
\textbf{87}, 054903 (2013).

\bibitem{Her13} C. Herold, M. Nahrgang, I. Mishustin, and M. Bleicher, Phys. Rev. C 
\textbf{87}, 014907 (2013).

\bibitem{Her14} C. Herold, M. Nahrgang, I. Mishustin, and M. Bleicher, Nucl. Phys. A 
\textbf{925}, 14 (2014).

\bibitem{Li17} F. Li and C. M. Ko, Phys. Rev. C 
\textbf{95}, 055203 (2017).

\bibitem{Sun22} K.J. Sun, W.H. Zhou, L.W. Chen, C.M. Ko, F. Li, R. Wang, and J. Xu, arXiv:2205.11010.

\bibitem{Nam61A} Y. Nambu and G. Jona-Lasinio, Phys. Rev. 
\textbf{122}, 345 (1961).

\bibitem{Nam61B} Y. Nambu and G. Jona-Lasinio, Phys. Rev. 
\textbf{124}, 246 (1961).

\bibitem{Bha17} A. Bhattacharyya, S.K. Ghosh, S. Maity, S. Raha, R. Ray, K. Saha, and S. Upadhaya, Phys. Rev. D 
\textbf{95}, 054005 (2017). 

\bibitem{Li19} Z.B. Li, K. Xu, X.Y. Wang, M. Huang, Eur. Phys. J. C 
\textbf{79},245 (2019).

\bibitem{Bha10} A. Bhattacharyya, P. Deb, S.K. Ghosh, and R. Ray, Phys. Rev. D 
\textbf{82}, 014021 (2010).

\bibitem{Sun21D} K.J. Sun, C.M. Ko, S. Cao, and F. Li, Phys. Rev. D 
\textbf{103}, 014006 (2021).

\bibitem{Yan24} Y.H. Yang, H. Liu, P.C. Chu, Nucl. Sci. Tech. 
\textbf{35}, 166 (2024).

\bibitem{Sun17} K.J. Sun, L.W. Chen, C. M. Ko, and Z. Xu, Phys. Lett. B
\textbf{774}, 103 (2017).

\bibitem{Hoo76} G. t’Hooft, Phys. Rev. D 
\textbf{14}, 3432 (1976); G. t’Hooft, Phys. Rev. D 
\textbf{18}, 2199 (E) (1978).

\bibitem{Mul95} H. Müller and B.D. Serot, Phys. Rev. C 
\textbf{52}, 2072 (1995).

\bibitem{Din15} H.T. Ding, F. Karsch, and S. Mukherjee, Int. J. Mod. Phys. E 
\textbf{24}, 1530007 (2015).

\bibitem{Abe09C} B.I. Abelev \textit{et al}., (STAR Collaboration), Phys. Rev. C 
\textbf{79}, 034909 (2009).

\bibitem{Baz14} A. Bazavov \textit{et al}., (HotQCD Collaboration), Phys. Rev. D 
\textbf{90}, 094503 (2014).

\bibitem{Ada17C} L. Adamczyk \textit{et al}., (STAR Collaboration), Phys. Rev. C 
\textbf{96} 044904 (2017).

\bibitem{Mot20} M. Motta, R. Stiele, W.M. Alberico, A. Beraudo, Eur. Phys. J. C 
\textbf{80}, 770 (2020).

\bibitem{Alb11} W.M. Alberico, A. Beraudo, A. De Pace, A. Molinari, M. Monteno, M. Nardi, F. Prino, Eur. Phys. J. C 
\textbf{71}, 1666 (2011).

\bibitem{Ran10} J. Randrup, Phys. Rev. C 
\textbf{82}, 034902 (2010).

\bibitem{Li16} F. Li and C.M. Ko, Phys. Rev. C 
\textbf{93}, 035205 (2016)

\bibitem{Sun21A} K.J. Sun, C.M. Ko, F. Li, J. Xu and L.W. Chen, Eur. Phys. J. A 
\textbf{57},313 (2021).

\bibitem{Mis99} I. N. Mishustin, Phys. Rev. Lett. \textbf{82}, 4779 (1999).

\bibitem{Rei07} W. Reisdorf \textit{et al}., (FOPI Collaboration), Nucl. Phys. A
\textbf{781}, 459 (2007).

\bibitem{Adl05}  S.S. Adler \textit{et al}., (PHENIX Collaboration), Phys. Rev. Lett. 
\textbf{94}, 122302 (2005).

\bibitem{Zha21} W. Zhao, K.J. Sun, C.M. Ko, and X. Luo, Phys. Lett. B
\textbf{820}, 136571 (2021).

\bibitem{Abo25} B.E. Aboona \textit{et al}., (STAR Collaboration), Phys. Rev. Lett.
\textbf{135}, 142301 (2025).
\end{thebibliography}
\end{document}